\documentclass[aps,prl,twocolumn,floatfix,showpacs,superscriptaddress]{revtex4}
\usepackage{graphicx}

\begin{document}
\title{Biaxial Strain in the Hexagonal Plane of MnAs Thin Films:
\\The Key to Stabilize Ferromagnetism to Higher Temperature}
\author{V. Garcia}\altaffiliation[Present address: Department of Materials Science, University of Cambridge, Cambridge, CB2 3QZ, UK - ] {e-mail: vg253@cam.ac.uk}
\affiliation{Institut des NanoSciences de Paris, INSP, Universit\'e
Pierre et Marie Curie-Paris 6, Universit\'e Denis Diderot-Paris 7,
CNRS UMR 7588, Campus Boucicaut, 140 rue de Lourmel, 75015 Paris,
France}
\author{Y. Sidis}
\affiliation{Laboratoire L\'eon Brillouin, CEA/CNRS, F-91191
Gif-sur-Yvette CEDEX, France}
\author{M. Marangolo}
\affiliation{Institut des NanoSciences de Paris, INSP, Universit\'e
Pierre et Marie Curie-Paris 6, Universit\'e Denis Diderot-Paris 7,
CNRS UMR 7588, Campus Boucicaut, 140 rue de Lourmel, 75015 Paris,
France}
\author{F. Vidal}
\affiliation{Institut des NanoSciences de Paris, INSP, Universit\'e
Pierre et Marie Curie-Paris 6, Universit\'e Denis Diderot-Paris 7,
CNRS UMR 7588, Campus Boucicaut, 140 rue de Lourmel, 75015 Paris,
France}
\author{M. Eddrief}
\affiliation{Institut des NanoSciences de Paris, INSP, Universit\'e
Pierre et Marie Curie-Paris 6, Universit\'e Denis Diderot-Paris 7,
CNRS UMR 7588, Campus Boucicaut, 140 rue de Lourmel, 75015 Paris,
France}
\author{P. Bourges}
\affiliation{Laboratoire L\'eon Brillouin, CEA/CNRS, F-91191
Gif-sur-Yvette CEDEX, France}
\author{F. Maccherozzi}
\affiliation{TASC laboratory, INFM–-CNR, in Area Science Park, S.S.
14, Km 163.5, I-34012 Trieste, Italy}
\author{F. Ott}
\affiliation{Laboratoire L\'eon Brillouin, CEA/CNRS, F-91191
Gif-sur-Yvette CEDEX, France}
\author{G. Panaccione}
\affiliation{TASC laboratory, INFM–-CNR, in Area Science Park, S.S.
14, Km 163.5, I-34012 Trieste, Italy}
\author{V. H. Etgens}
\affiliation{Institut des NanoSciences de Paris, INSP, Universit\'e
Pierre et Marie Curie-Paris 6, Universit\'e Denis Diderot-Paris 7,
CNRS UMR 7588, Campus Boucicaut, 140 rue de Lourmel, 75015 Paris,
France}

\begin{abstract}
The $\alpha-\beta$ magneto-structural phase transition in
MnAs/GaAs(111) epilayers is investigated by elastic neutron
scattering. The in-plane parameter of MnAs remains almost constant
with temperature from 100 K to 420 K, following the thermal
evolution of the GaAs substrate. This induces a temperature
dependent biaxial strain that is responsible for an $\alpha-\beta$
phase coexistence and, more important, for the stabilization of the
ferromagnetic $\alpha$-phase at higher temperature than in bulk. We
explain the premature appearance of the $\beta$-phase at 275 K and
the persistence of the ferromagnetic $\alpha$-phase up to 350 K with
thermodynamical arguments based on the MnAs phase diagram. It
results that the biaxial strain in the hexagonal plane is the key
parameter to extend the ferromagnetic phase well over room
temperature.
\end{abstract}
\pacs{75.70.Ak, 61.50.Ks, 68.60.-p}

\date{\today}
\maketitle

The first-order phase transition near room temperature and the
magnetic properties of bulk manganese arsenide (MnAs) have been
intensively investigated in the last century \cite{BeanRodbell1962,
MenyukPR1969, PytlikJMMM1985}. Recently, the possibility of
epitaxial growth of MnAs thin films on standard semiconductors such
as GaAs has renewed the interest of MnAs for spintronics related
research \cite{TanakaAPL1994}. Indeed, high quality epilayers of
MnAs can be grown on GaAs substrates with a very low reactivity
between the ferromagnet and the semiconductor \cite{TanakaAPL1994,
SchippanJVacSciTech1999, MattosoPRB2004, GarciaPRB2006}. Also,
electrical spin injection in GaAs has been recently demonstrated
with MnAs \cite{StephensPRL2004} and spin-polarized tunneling
performed with MnAs/GaAs/MnAs junctions \cite{GarciaPRB2005}. Even
if MnAs has a great advantage over common transition metals in terms
of reactivity with GaAs, its Curie temperature, associated to the
first-order ferromagnetic-paramagnetic phase transition, is only
slightly above room temperature. A great challenge is to understand
this phase transition to maintain ferromagnetism in MnAs to higher
temperature.

Bulk manganese arsenide (MnAs) is a room temperature ferromagnetic
and metallic compound up to T$_{C}$=313 K where the first-order
phase transition from hexagonal ($\alpha$-phase, NiAs type) to
orthorhombic ($\beta$-phase, MnP type) is accompanied by a
ferromagnetic-paramagnetic transition \cite{WillisRooksby1954,
MenyukPR1969}. A large volume contraction of about 2$\%$ is observed
at T$_{C}$ and this contraction occurs mainly in the hexagonal basal
plane. An hydrostatic pressure of a few kbars can however stabilize
the $\beta$-phase over the $\alpha$-phase below T$_{C}$
\cite{MenyukPR1969} [see Fig. \ref{Fig3}(a)].

\begin{figure}
\centering
\includegraphics[scale=0.8]{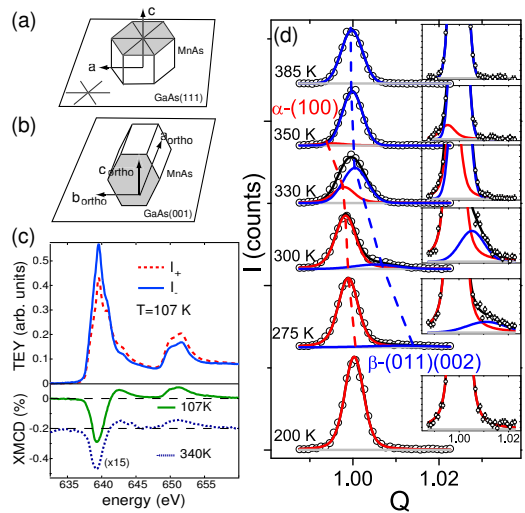}
\caption{(color online) Schemes of MnAs epitaxy (a) on GaAs(111)B
and (b) on GaAs(001) with an orthorhombically distorted unit cell.
(c) Mn L$_{2,3}$ XAS spectra for antiparallel (dotted) and parallel
(continuous) helicity and magnetization, at 107 K; difference
spectrum (XMCD) (I$_{+}$+I$_{-}$) normalized to the L$_{3}$ peak of
the summed spectra (I$_{+}$-I$_{-}$) and scaled for photon
polarization (70$\%$) and incidence angle (45$^{\circ}$), at 107 K
(line) and 340 K (dot). Sample magnetization lies in plane. The 340
K spectrum is multiplied by 15 for clarity. The lineshape remains
unchanged at high temperature. (d) Temperature dependence of the
neutron diffraction patterns of MnAs (100 nm)/GaAs(111)B for
in-plane (100)-$\alpha$ and corresponding (011)-(002)-$\beta$
reflections. The dashed lines following the maxima of each Bragg
peak are only guides for the eyes. Zooms of the patterns are shown
in inset.} \label{Fig1}
\end{figure}

Interestingly, in the case of MnAs epilayers grown on GaAs
substrates, the epitaxial strain disturbs the first-order phase
transition. MnAs epilayers grown on GaAs(100) substrates have been
deeply investigated. The epitaxy is sketched in Figure \ref{Fig1}(b)
with the hexagonal $c$ axis aligned in the film plane. Kaganer
\emph{et al.} \cite{KaganerPRL2000} have shown that the epitaxial
strain leads to the $\alpha-\beta$ phase coexistence to minimize the
elastic energy. Strain dependent magnetic properties were analyzed
by Das \emph{et al.} \cite{DasPRL2003} and Iikawa \emph{et al.}
\cite{IikawaPRL2005} in the case of MnAs epitaxied on GaAs(001). The
unit cell of the ferromagnetic $\alpha$-phase was found to be
orthorhombically distorted, as the hexagonal plane of MnAs lies out
of the substrate surface. This temperature dependent structural
modification induces the early appearance of the paramagnetic
$\beta$-phase at T$_{C}^{(1)}$=273 K \cite{AdrianoAPL2006}.
Minimization of elastic energy makes the two phases coexist up to
T$_{C}^{(2)}$=315 K which is almost the same transition temperature
as for the bulk material.

In order to probe the effect of a biaxial strain on this transition,
we have grown 100 nm thick MnAs(001) thin films on GaAs(111)B
substrates. The films display a single epitaxy with the hexagonal
$c$ axis along the growth direction and the hexagonal plane over the
hexagonal (111) surface of the GaAs substrate [Fig. \ref{Fig1}(a)]
\cite{MattosoPRB2004}. Magnetic measurements reveal a bulk like
saturation magnetization (900-950 emu/cm$^3$) at low temperature
with high remanence and low coercive fields \cite{MattosoPRB2004}.
The critical temperature is significantly enhanced with a Curie
point (deduced from $\partial M/\partial T$ curve) of 335 K and a
magnetization is detected up to 350 K. X-ray magnetic circular
dichroism (XMCD) spectra collected at the APE beamline (Elettra
synchrotron, Trieste) attest that ferromagnetism is stable up to 340
K. No significant modification of the $\alpha$-pure XMCD lineshape
is detected from 100 K to 340 K [Fig. \ref{Fig1}(c)]: this implies
that persistent ferromagnetism can be ascribed to the
$\alpha$-phase. A detailed discussion about absorption spectra is
beyond the scope of this letter and will be published elsewhere.

In this letter, we show magneto-structural characterizations of
MnAs(001) thin films by neutron diffraction experiments in a wide
range of temperature (100 K - 420 K). The decisive advantage of
neutron scattering is that it enables us to transmit through the
substrate and measure directly the in-plane and growth-axis lattice
parameters from selected Bragg reflections. The films display large
in-plane deformations compared to the bulk, inducing an
$\alpha-\beta$ phase coexistence from 275 K to 350 K. The
out-of-plane parameter has an almost bulk-like behavior. The mean
in-plane parameter 
%\footnote{The mean in-plane parameter is defined
%as $a_{mean}(T)=a_{\alpha}(T)\times
%frac(\alpha)(T)+a_{\beta}(T)\times [1-frac(\alpha)(T)]$.} 
is almost constant from 100 K to 420 K and follows the expansion coefficient
of the GaAs substrate. We estimated the strain induced in each phase
from the measured deformations in the plane. Considering the biaxial
strain equivalent to an hydrostatic pressure, we have succeeded to
explain the early appearance of the $\beta$-phase (T$_{C}^{(1)}$=275
K) and the high temperature preservation of the $\alpha$-phase
(T$_{C}^{(2)}$=350 K). Finally, we conclude that the stability of
the ferromagnetic phase is strongly dependent on the epitaxial
strain in the basal plane.

Elastic neutron scattering measurements were performed on the
triple-axis spectrometer 4F1, installed on a cold neutron beam at
the reactor Orph\'ee in Saclay (France). A monochromatic neutron
beam was obtained with a double monochromator, made of pyrolytic
graphite in (002) reflection [PG(002)], and filtered by cold
beryllium to eliminate higher-order contamination. The diffracted
beam was analyzed by a PG(002) crystal, with 40$^{'}$ collimators on
each side. The initial neutron wave vector was set to
$k_{i}=1.2$~\AA$^{-1}$. A 100 nm thick MnAs thin film was grown by
molecular beam epitaxy on a GaAs(111)B substrate as described
elsewhere \cite{GarciaPRB2006}, and capped \emph{in situ} with a
thin gold layer (5 nm) to prevent oxidation. The sample with a
surface of about 2 cm$^{2}$ was mounted in an aluminum can
containing helium exchange gas, and fixed on the cold finger of a
closed cycle refrigerator, operating from 100 K to 420 K. The sample
was aligned so that (100), (101) and (002) Bragg reflections of the
hexagonal phase were accessible. We monitored the temperature
dependence of the in-plane and out-of-plane parameters, from 100 to
420 K; each temperature point was preceded, after temperature
stabilization ($\approx$ 30 min), by a rocking scan on GaAs(111)
reflection.

The triple-axis spectrometer resolution \cite{Neutrons} was adjusted
from the (111) and (1$\overline{1}$0) reflections of the GaAs
substrate which we assume to be a perfect crystal. We find that in
both in-plane and out-of-plane directions, the experimental
resolution is defined by a Gaussian function with a full width half
maximum (FWHM) of $\sim0.01$~\AA$^{-1}$. The analysis of the neutron
data was performed self-consistently around three different wave
vectors: (100), (101) and (002) of the hexagonal phase. Bragg peaks
were fitted with Gaussian functions convoluted with the experimental
resolution function. The intensities, widths and positions of the
Gaussian functions are free parameters in our analysis. Out-of-plane
measurements were performed along $\alpha$-(002) [and $\beta$-(200)]
peaks. They show an almost constant intrinsic width from 100 to 420
K, corresponding to a correlation length of 80 nm perpendicular to
the film, in good agreement with the film thickness (100 nm).
In-plane measurements reveal an in-plane correlation length of 70 to
80 nm. Three temperature cycles (100 K$\rightarrow$420
K$\rightarrow$100 K) were performed showing very reproducible
structural parameters without any hysteresis between heating and
cooling.

Figure \ref{Fig1}(d) displays the temperature dependence of the
radial scan along $\alpha$-(100) including the $\beta$-(011)(002)
doublet. The $\beta$-phase is found to nucleate around 275 K and the
$\alpha$-phase is present up to 350 K. In this temperature range,
the neutron data cannot be described by a single peak anymore. Owing
to their intrinsic widths, the two Gaussian functions associated
with the Bragg peaks of the $\alpha$- and $\beta$-phases heavily
overlap, but the peak positions can still be determined accurately
(see zooms in Fig. \ref{Fig1}(d)).

\begin{figure}
\centering
\includegraphics[scale=0.7]{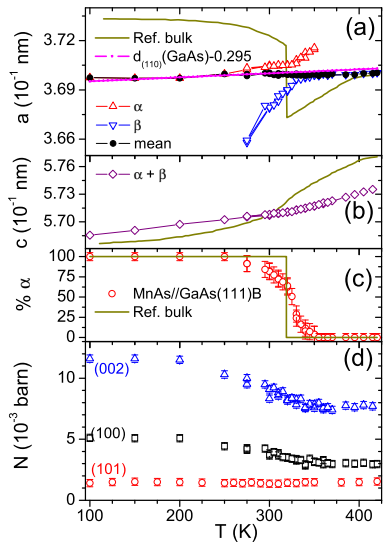}
\caption{(color online) Temperature dependence of structural
parameters of a 100 nm thin film of MnAs grown on GaAs(111)B
substrate: (a) in-plane parameter for both phases deduced from
neutron diffraction experiments on MnAs(100), (b) out-of-plane
parameter deduced from scans along MnAs(002). The in-plane and
out-of-plane parameters of bulk MnAs (adapted from Ref.
\protect\onlinecite{SuzukiJPhysSocJpn1982}) are added. (c) $\alpha$-phase
fraction deduced from (d) the integrated intensity of the spectra
along MnAs(100), (101) and (002). The mean in-plane parameter
calculated from this $\alpha$-phase fraction, is added in (a)
together with the evolution of GaAs parameter.} \label{Fig2}
\end{figure}

\begin{figure}
\centering
\includegraphics[scale=0.5]{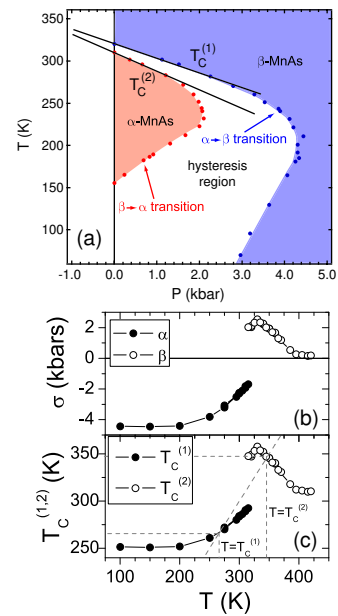}
\caption{(color online) (a) Temperature-pressure phase diagram of
bulk MnAs adapted from Ref. \protect\cite{MenyukPR1969}. Pressure
dependencies of the transition temperatures of -15 K/kbar and -20
K/kbar are deduced for T$_{C}^{(1)}$ and T$_{C}^{(2)}$ respectively.
(b) Calculated in-plane strain deduced from the lattice deformation
compared to bulk values in Fig. \ref{Fig2}(a). (c) Calculated
critical temperature considering the in-plane strain equivalent to
pressure: the critical points are deduced from the intersection
between the experimental points and the straight line T=T$_{C}$.}
\label{Fig3}
\end{figure}

The temperature evolution of the lattice parameters of MnAs
epitaxied on GaAs(111)B are compared to bulk MnAs values
\cite{SuzukiJPhysSocJpn1982} in Figure \ref{Fig2}. Strong
epitaxy-induced modifications of MnAs thin films parameters are
observed. At low temperatures the in-plane $\alpha$-phase MnAs unit
cell is compressed as compared to the bulk; strain remains stable up
to the appearance of the $\beta$-phase at 275 K [Fig.
\ref{Fig2}(a)]. As already observed by Mattoso \emph{et al.}
\cite{MattosoPRB2004}, the $\beta$-phase displays a strong
temperature dependence of its lattice parameter in the coexistence
temperature range. The nucleation of the $\beta$-phase allows the
$\alpha$-phase in-plane parameter to relax towards its bulk value. A
tensile strain is observed in the film plane for the pure
$\beta$-phase and it progressively disappears at higher temperature
[Fig. \ref{Fig2}(a)].

Around each of the three different wave vectors of the hexagonal
phase (100), (101) and (002), we further computed the neutron
scattering cross-section associated with Bragg reflections of the
paramagnetic $\beta$-phase and ferromagnetic $\alpha$-phase,
assuming a temperature independent magnetization of 3.6 $\mu_{B}$
per Mn \cite{Squires1978}. This intensity is proportional to the
volume fraction of each phase in the film. Thus for any wave vector,
the temperature dependence of the total norm of the measured
$\alpha$- and $\beta$-peaks allows us to determine the volume
fraction of each phase. Figure \ref{Fig2}(c) shows the temperature
dependence of the volume fraction of the $\alpha$-phase, deduced
from the temperature dependencies of the total norm of the peak(s)
measured around three different wave vectors of the hexagonal phase
(100), (101) and (002) [Fig. \ref{Fig2}(d)].

%The temperature evolution of the volumic fraction of the
%$\alpha$-phase is shown in Figure \ref{Fig2}(c). It was obtained by
%calculating the integrated areas of decomposed $\alpha$-(100) and
%$\beta$-(011)(002) peaks. Since the neutron cross-section is
%affected also by a magnetic contribution, we obtained the
%$\alpha$-phase fraction (i) by evaluating the magnetization
%comparing the integrated area of the ferromagnetic and paramagnetic
%peaks at 100 K and 400 K respectively, and (ii) by considering a
%constant atomic magnetic moment up to 350 K. The obtained fraction
%curve fits very well with the data obtained by Mattoso \emph{et al.}
%\cite{MattosoPRB2004} by X-ray scattering, corroborating this
%procedure.

The conservation of the thin film lateral size
leads to a thermal evolution that is not as free as it is possible
for a bulk single crystal. Since our films are continuous and
epitaxied, the temperature evolution of the mean lateral lattice
spacing in the film should follow the thermal lattice expansion
coefficient of the GaAs substrate. This condition is fulfilled as
shown in Figure \ref{Fig2}(a), where we compare the measured
in-plane thermal evolution of GaAs ($d_{110}$) with the mean lateral
lattice spacing of MnAs, obtained by considering the evaluated
volumic fraction of each phase at a given temperature [Fig.
\ref{Fig2}(c)]. In other words, MnAs thin films are almost fully
relaxed at the growth temperature (T$\cong$500 K) because the large
lattice mismatch (7-8$\%$) is released by dislocations but, when
temperature is reduced no more dislocations can be generated and the
"frozen" films undergo two lateral size preserving phase
transitions.

The out-of-plane parameter evolution with temperature is close to
that of bulk MnAs [see Fig. \ref{Fig2}(b)]. The temperature
evolution of the in-plane parameters of the thin films can be
quantitatively compared to those of bulk MnAs to evaluate the strain
incorporated in the material. Cooling the sample from growth
temperature, an in-plane progressive tensile strain is introduced in
MnAs. In the following, we put forward the hypothesis that the
biaxial strain in the hexagonal plane of MnAs is the leading
parameter for the pressure dependencies of T$_{C}^{(1)}$ and
T$_{C}^{(2)}$ reported by Menyuk \emph{et al.} [Fig. \ref{Fig3}(a)]
\cite{MenyukPR1969}.

Consequently, the persistence of the $\alpha$-phase at higher
temperature (T$_{C}^{(2)}$) in MnAs/GaAs(111)B thin films and the
early nucleation of the $\beta$-phase at lower temperature
(T$_{C}^{(1)}$) can be understood by thermodynamical considerations.
For example, at 360 K an in-plane lattice expansion of +0.36$\%$ is observed
in the $\beta$-phase corresponding to a tensile strain of 1.7 kbars. Also,
at 250 K an in-plane lattice compression of -0.82$\%$ in the $\alpha$-phase
corresponding to a compressive strain of 3.8 kbars
\footnote{The in-plane strain $\sigma_{1}$ is calculated from the
lattice expansion $\varepsilon_{1}$ along [110] axis of MnAs
considering planar constraints ($\sigma_{3}$=0 along [001] axis of
MnAs):
$\sigma_{1}=(c_{11}+c_{12}-2c_{13}^{2}/c_{33})\varepsilon_{1}$. The
elastic constants $c_{ij}$ were taken from M. D\"{o}rfler and K.
B\"{a}rner, Phys. Status Solidi (a) \textbf{17}, 141 (1973).}. From
Fig. \ref{Fig3}(a), the negative slope of T$_{C}^{(2)}$ with
pressure ($\partial T_{C}^{(2)}/\partial P$=-20 K/kbar with
T$_{C,0}^{(2)}$=307 K, \cite{MenyukPR1969}) is coherent with the
enhancement of T$_{C}^{(2)}$ with tensile strain (negative pressure). In
parallel, the compressive strain in the $\alpha$-phase lowers
T$_{C}^{(1)}$ ($\partial T_{C}^{(1)}/\partial P$=-15 K/kbar with
T$_{C,0}^{(1)}$=313 K, \cite{MenyukPR1969}).

On the other hand, since MnAs thin films grown on GaAs(001) are
virtually strain-free in the basal plane of the $\beta$-phase above
the phase transition \cite{DasPRL2003}, the critical temperature
(T$_{C}^{(2)}$) is not enhanced compared to the bulk material.

Going through details, we can estimate the two critical points where
the phase transition should proceed given the strain incorporated by
calculating the in-plane strain in both phases for each experimental
point [Fig. \ref{Fig3}(b)]. We deduced a first critical point at
T$_{C}^{(1)}$=265 K corresponding to the onset of the $\alpha$ to
$\beta$ phase transition and a second one at T$_{C}^{(2)}$=347 K
corresponding to the onset of the $\beta$ to $\alpha$ phase
transition [Fig. \ref{Fig3}(c)]. We observe a very good agreement
between the calculated critical point and the experimental
$\alpha$-$\beta$ phase coexistence observed between 275 K and 350 K.
When the sample is heated, from the first critical point
T$_{C}^{(1)}$=265 K, the $\beta$-phase nucleation induces a partial
relaxation of the compressive strain in the ferromagnetic phase and
the transition temperature is increased [Fig. \ref{Fig3}(c)]. This
behavior explains why in the epitaxial system, a phase coexistence
is observed instead of a total $\alpha$-$\beta$ phase transition.

The general agreement between the phase diagram of bulk MnAs and the
pressure induced by the biaxial strain in MnAs(001) thin films leads
to the conclusion that this strain is the key parameter for the
$\alpha$-$\beta$ phase transition. We demonstrate that biaxial
strain in the hexagonal plane of MnAs thin films can significantly
enhance the stability of the ferromagnetic phase with temperature.
We anticipate that larger T$_{C}^{(2)}$ critical temperature may be
obtained by increasing the tensile strain in the film plane. This
could be achieved by applying external strain to the substrate
similarly to the experiments performed by Iikawa \emph{et al.}
\cite{IikawaPRL2005}. For example, if we apply a tensile strain of
0.5$\%$ in the film plane, it would be possible to increase the
critical temperature to T$_{C}^{(2)}$=373 K \footnote{Applying
biaxial tensile strain to the substrate will induce a positive shift
of the mean in-plane parameter in Fig. \ref{Fig2}(a). T$_{C}^{(2)}$
is calculated from the strain induced in the $\beta$-phase using
Fig. \ref{Fig3}(a) and Ref. 18.}. Another way to enhance
ferromagnetism to higher temperature is to grow non-relaxed
MnAs(001) thin films on a small mismatched (111) cubic substrate
where the MnAs hexagonal basal plane would remain under tensile
strain.

%\vspace{0.5cm}

%This research was supported by a French Program of Nanosciences and
%Nanotechnology (PNANO).

%\bibliography{bibliothese}

\end {document}